\documentclass[12pt]{iopart}
\usepackage{iopams}
\usepackage{setstack}
\usepackage{amsthm}
\usepackage{listings}
\usepackage{enumerate}
\usepackage{latexsym}
\usepackage[dvips]{graphicx}
\usepackage{psfrag}
\usepackage{bm}
\usepackage{framed}

\newcommand{\beq}{\begin{equation}}
\newcommand{\eneq}{\end{equation}}

\newcommand{\braket}[2]{\left\langle #1 | #2 \right\rangle}
\newcommand{\bra}[1]{\left\langle#1\right|}
\newcommand{\ket}[1]{\left|#1\right\rangle}

\def\binom#1#2{{#1\choose#2}}

\newcommand{\of}[1]{\!\left(#1\right)}

\input{epsf}

\begin{document}

\tolerance 10000

\title{Hierarchical structure in the orbital entanglement spectrum in  Fractional Quantum Hall  systems}

\author{A. Sterdyniak$^{1}$, B.A. Bernevig$^{2}$, N. Regnault$^{1}$ and F. D. M. Haldane$^2$ }

\address{$^1$ Laboratoire Pierre Aigrain, ENS and CNRS, 24 rue Lhomond, 75005 Paris, France}
\address{$^2$ Department of Physics,
  Princeton University, Princeton, NJ 08544}

\ead{}

\begin{abstract}
We investigate the non-universal part of the orbital entanglement spectrum (OES)  of the $\nu = 1/3$
fractional quantum Hall effect (FQH) ground-state with Coulomb interactions. The non-universal part of the spectrum is the part that is missing
in the  Laughlin model state OES whose level counting is completely determined by its topological order. We find that the OES levels of the Coulomb interaction ground-state  are organized in a hierarchical structure that mimic the excitation-energy structure of the model pseudopotential Hamiltonian which has a Laughlin ground state. These structures can be accurately modeled using Jain's ``composite fermion'' quasihole-quasiparticle excitation wavefunctions. To emphasize the connection between the entanglement spectrum and the energy spectrum, we also consider the thermodynamical OES of the model pseudopotential Hamiltonian at finite temperature. The observed good match between the thermodynamical OES and the Coulomb OES  suggests a relation between the entanglement gap and the true energy gap.
\end{abstract}
%\submitto{}
%\tableofcontents
\maketitle

\section{Introduction}

Most condensed matter phases (states of matter) can be characterized using local order parameters. However, in some  cases, such as  systems with topological order, it is not possible to obtain a full characterization of a state using only local operators. It has been proposed that non-local measurements borrowed from quantum information theory\cite{RevModPhys.80.517}, such as entanglement,  can provide new insights into topological phases  phases. The most commonly-used quantifier of entanglement, the Von Neumann  entanglement entropy, measures the entanglement between two blocks of the system. In systems exhibiting the  Fractional Quantum Hall (FQH) effect, the first topological phases to be experimentally realized, such an entanglement measure providing a single number does not provide a unique characterization of the many  possible states that can occur. A few  years ago, it was suggested that generalizing the entanglement entropy to the entanglement spectrum \cite{li-08prl010504}, could  reveal much more of the physical properties of a FQH state.

The  notion of entanglement spectrum has now been applied to many different systems: Quantum Hall mono-layers \cite{li-08prl010504,regnault-09prl016801, thomale-10pr180502, lauchli-10prl156404, lauchli-NJP-1367-2630,bergholtz-arXiv1006.3658B,PhysRevLett.106.100405,rodriguez-arXiv1007.5356R,papic-arXiv1008.5087P,kargarian-82prb085106,hermanns-1009arXiv4199H,PhysRevB.80.201303,2011arXiv1103.5437Q, 2011arXiv1103.0772Z}, Quantum Hall bilayers\cite{PhysRevB.83.115322, thomale-2010arXiv1010.4837T}, quantum spin systems \cite{thomale-105prl116805, poilblanc-105prl077202, turner-2010arXiv1008.4346T, fidkowski-81prb134509, yao-105prl080501, pollmann-njp-1367-2630, 2010PhRvB..81f4439P, 2008PhRvA..78c2329C, 2010arXiv1010.4508H, PhysRevB.83.045110, 2011arXiv1104.2544S, 2010arXiv1011.2147P, 2010arXiv1002.2931F,2011arXiv1104.1139H, 2011arXiv1103.3427C}, superconductor\cite{bray-ali-80prb180504,2011arXiv1105.4808D}, topological insulators \cite{prodan-105prl115501,fidkowski-104prl130502,2010PhRvB..82x1102T} and ultra-cold gases \cite{PhysRevA.83.013620, 2011arXiv1104.5157D}. For a system described by a density matrix $\rho$, the orbital entanglement spectrum (OES) is defined using an angular-momentum orbital  decomposition that cuts the system into two regions $A$ and $B$. Such an orbital cut mimics a geometrical cut only when the orbitals are localized in space. The reduced density matrix $\rho_A$ is obtained by tracing out the $B$ subsystem degrees of freedom, which yields $\rho_A=\Tr_B \rho$. As the eigenvalues of $\rho_A$  are non-negative, one can write $\rho_A=\exp(- \mathcal{H})$, thus introducing a fictitious Hamiltonian $\mathcal{H}$, whose spectrum is the OES.

For several FQH states, the count of the low-lying states in the entanglement spectrum was numerically shown to match that of the state's edge-theory. Moreover, for a realistic Hamiltonian ground state that has a large overlap with a FQH model wavefunction, the OES exhibits a low-lying branch with the same entanglement state-count as the model wavefunction, but now accompanied by higher  entanglement energy-levels which were previously not believed to provide useful information about the system.  In this work, we show that the higher energy  levels in the Coulomb OES are organized into branches whose structure can be related to virtual particle-hole excitations  that dress the simpler entanglement spectrum of the model ground-state that just characterizes (in its purest form) the topological order of the FQH state.

This paper is organized as follows. In Section \ref{section_ESinFQHE}, we introduce the sphere geometry, present the concept of the entanglement spectrum for the fractional quantum Hall effect, and summarize the main results that have already been obtained in the literature. In Section \ref{section_JainCF}, we introduce Jain's composite fermion wavefunctions and their neutral excitations. In Section \ref{section_laughlinCoulombES}, we use the composite fermion construction to interpolate between Laughlin state and Coulomb interaction ground state entanglement spectrum.  This interpolation explains the hierarchical structure observed in the OES. In Section \ref{section_thermalES}, we investigate the behavior of the entanglement spectrum at finite temperature and, based on it, conjecture a relation between the energy gap and the entanglement gap.  In the final section \ref{section_discussion}, we present a discussion of these results.

\section{Entanglement spectrum of  fractional quantum Hall States on the sphere}
\label{section_ESinFQHE}

We will consider a system of $N$ particles moving on the surface of a sphere $(\theta, \phi)$ through which  $N_{\Phi}$ magnetic flux quanta pass. The radius of the sphere is equal to $R=\sqrt{N_{\Phi}/2}$. In the lowest Landau level (LLL), the one-particle orbitals can be expressed as
$$
\psi_l(u,v)= \left[\frac{N_{\Phi}+1}{4\pi}\binom{N_{\Phi}}{l}\right]^{1/2}(-1)^{N_{\Phi}-l}v^{N_{\Phi}-l}u^{l}
$$
where $\binom{n}{k}$ is the binomial coefficient, $u=\cos(\theta/2)\rme^{\rmi \phi/2}$ and $v=\sin(\theta/2)\rme^{-\rmi \phi/2}$ are the spinor variables. These orbitals are eigenstates of $L_Z$, the $z$-component of the angular momentum, with eigenvalues given by $l- N_{\Phi}/2$, where $l$ ranges from $0$ to $N_{\Phi}$. The orbitals $\psi_l$ form an approximate real-space partition of the sphere into rings: the north pole (resp. south) corresponds to $l=N_{\Phi}$ (resp. $l=0$).

Generic fermionic (bosonic) many-body wave functions of $N$ particles and total azimuthal angular momentum $L_z^{tot}$ can be expressed as linear combinations of Fock states in the occupancy basis of the single particle orbitals. Each Fock state can be labeled either by $\lambda$, a partition of  $L_z^{tot}$ into $N$ components, or the occupation number configuration $n(\lambda)=\{n_l(\lambda), l = N_{\Phi},...,0\}$, where $n_l(\lambda)$  is the number of times $l$ appears in $\lambda$. We define ``\textit{squeezing}'' as  a two-particle operation on partitions that moves a particle from orbital $l_1$ (resp. $l_2$) to orbital $l'_1$ (resp. $l'_2$) such that $l_1+l_2=l'_1+l'_2$ and $l_1 < l'_1 \leq l'_2<l_2$ for bosons or $l_1 < l'_1 < l'_2<l_2$ for fermions. Squeezing defines a partial ordering on partitions: if a partition $\mu$ can be obtained from a partition $\lambda$ using successive squeezing operations, the partition $\lambda$ is said to \textit{dominate} the partition $\mu$ ($\lambda > \mu$). Certain model wave functions have a {\it root} partition $\lambda_0$: in their expansion on the Fock states, only partitions dominated by $\lambda_0$ can have a non-zero weight. For instance, this is the case in the $1/m$ Laughlin state, in which the occupation number configuration of the root partition is given by $n(\lambda_0)=\{10^{m-1}10^{m-1}1\dots \}$, where $0^{m-1}$ denotes $m-1$ consecutive empty orbitals. This root partition is ``$(1,m)$-\textit{admissible}'' -  it obeys a generalized Pauli principle which does not allow  more than one particle  to occupy $m$ consecutive orbitals.

The orbital entanglement spectrum (OES) is obtained by cutting the sphere into two parts $A$ and $B$. Part  $A$  contains the $l_A$ first orbitals from the north pole while part  $B$  contains the $l_B=N_{\Phi}+1 -l_A$ remaining orbitals. Due to the localized nature of the Landau level (LL) orbitals, this cut is a reasonable approximation to a spatial one \cite{haque-07prl060401}. Tracing over the $l_B$ orbitals gives a reduced density matrix $\rho_A$ with a  block-diagonal structure. Each block is characterized by two quantum numbers, $(L_{z,A},N_A)$:  the $z$-component of the angular momentum for the part-$A$ orbitals and the number of particles in part $A$  are the only  symmetry-generators present after the partial trace (the full $\vec{L}$ symmetry of the original state is lost). For the sphere geometry, the OES is the plot of the negative logarithm $\xi_i$ of $\rho_A$ eigenvalues as a function of $L_{z,A}$ for a fixed $N_A$. The $\xi_i$'s are called ``entanglement (pseudo)energies''.

For numerical efficiency we use an algorithm (described in Table \ref{algo}) that computes each block of the reduced density matrix independently, as  for a given pair of quantum numbers of the $A$ region, $(L_{z,A},N_A)$, the only $B$-region subspace to consider is that defined by $(L_{z,B}=L_z-L_{z,A},N_B=N-N_A)$.  When we consider a system that is not in a pure state, this procedure is repeated for all states present in the density matrix. The reduced density matrix thus obtained is then diagonalized using standard full diagonalization techniques.

\begin{table}
\begin{framed}
For each basis states $\ket{\Psi^B_i} \in B$ ($i$ denotes the position in the basis)\\
\hspace*{0.5cm}for each basis states $\ket{\Psi^A_j} \in A$\\
\hspace*{1cm}create the  basis state $\ket{\Psi}=\ket{\Psi^A_j}\otimes\ket{\Psi^B_i}$\\
\hspace*{1cm}find $k$, the index of $\ket{\Psi}$ in the basis describing $A\otimes B$ and the coefficient $c_k$ of\\\hspace*{1cm} in the full state $\ket{\Phi}$, including sign from reordering if needed (fermionic case)\\
\hspace*{0.5cm}for each index $k$ in the basis of $A$\\
\hspace*{1cm}for each index $p$ in the basis of $A$ with $p \geq k$\\
\hspace*{1.5cm}add to the matrix element $\rho(k,p)$: $c_k*c_p$
\end{framed}
\caption{\label{algo}: Algorithm that computes a given block $A$ of the reduced density matrix $\rho=\ket{\Phi}\bra{\Phi}$  defined by the quantum numbers $(L_{z,A},N_A)$ of the reduced density matrix $\rho$ for the state $\ket{\Phi}$. The complementary space $B$ is defined by $(L_{z,B}=L_z-L_{z,A},N_B=N-N_A)$}
\end{table}

\begin{figure}[htb]
\includegraphics[width=7.5 cm]{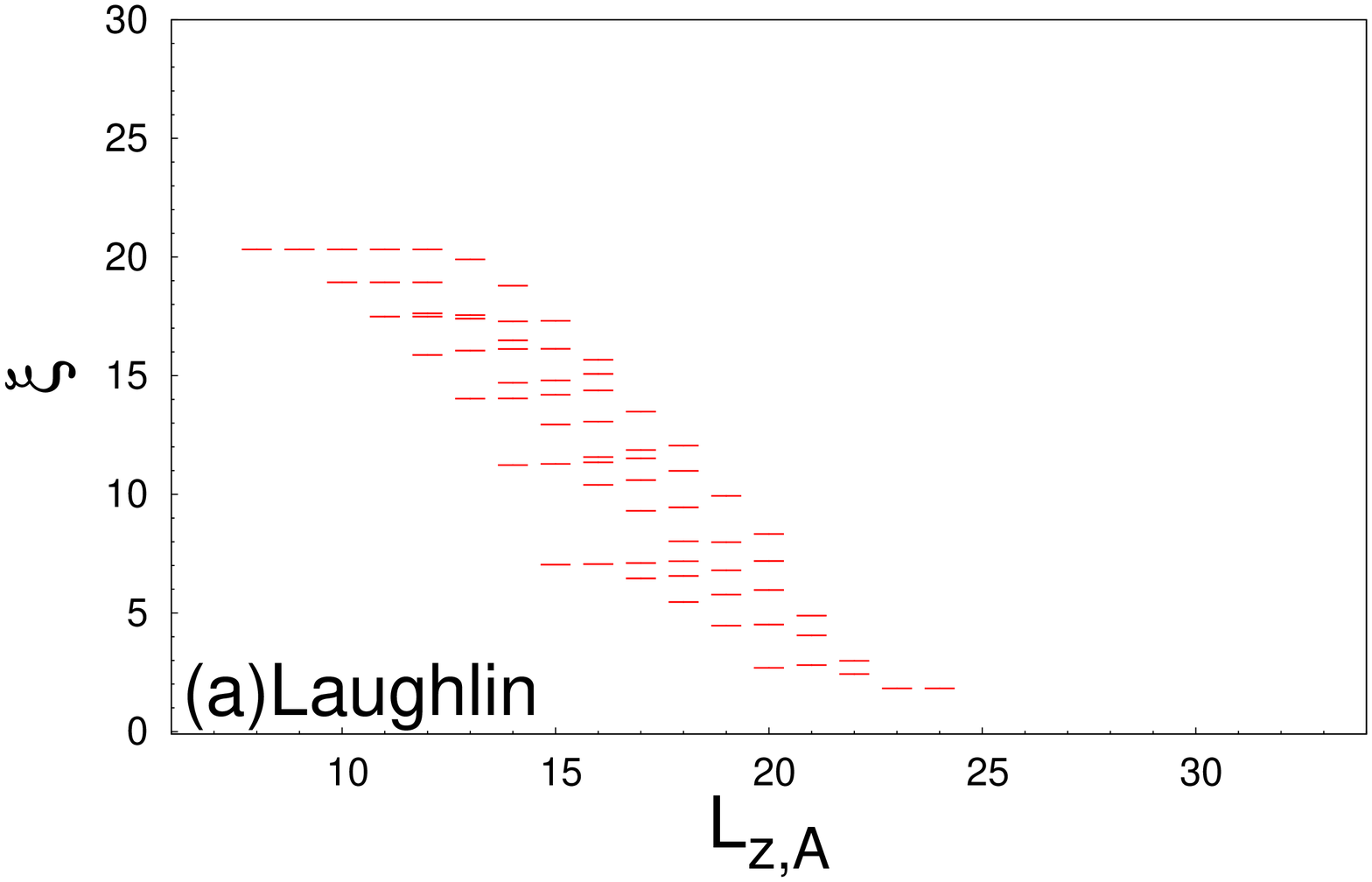}
\includegraphics[width=7.5 cm]{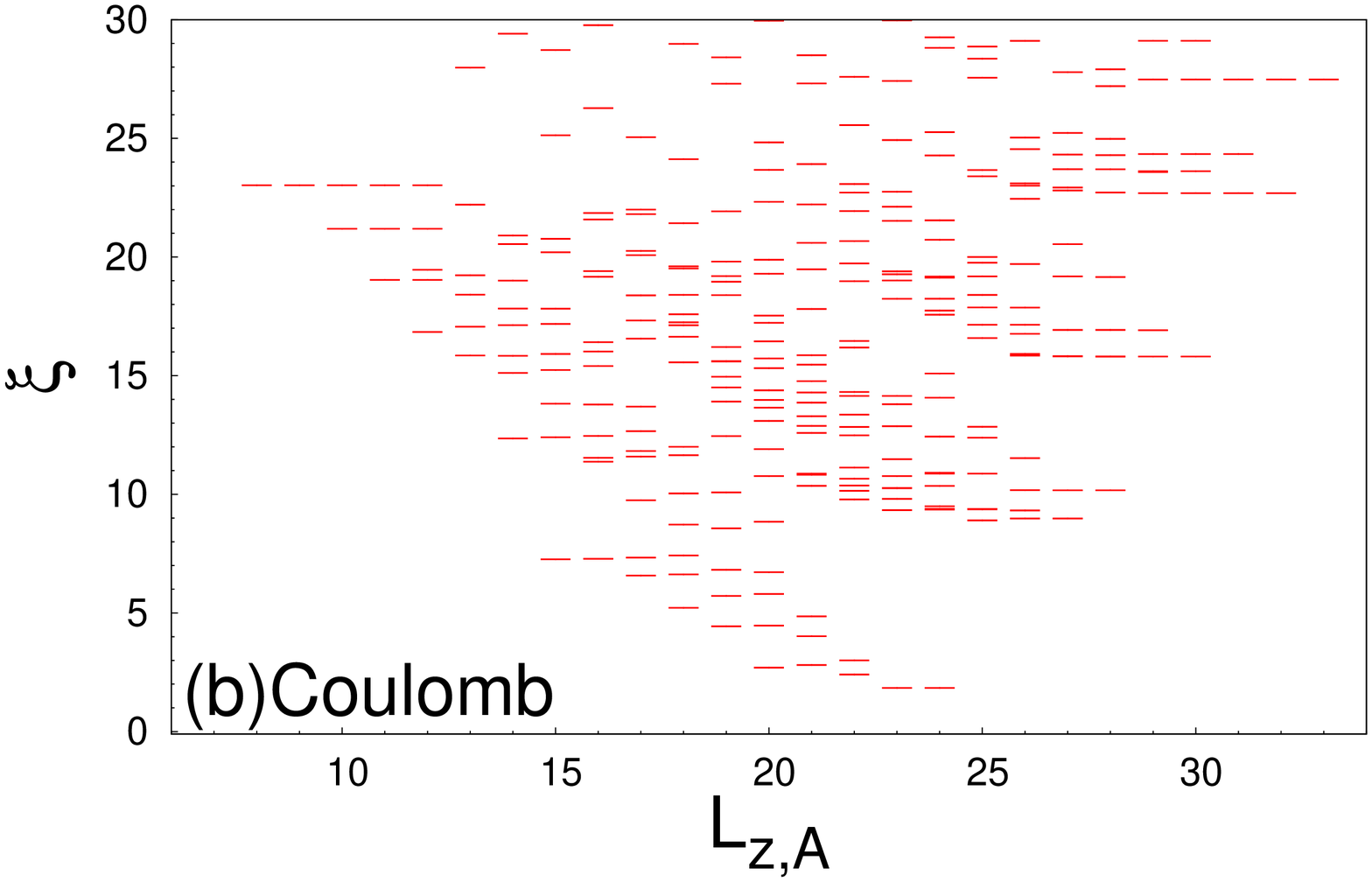}
\caption{\label{figure1}Orbital entanglement spectrum for the Laughlin state, (a), and for the ground state of the Coulomb interaction, (b) for $N=8$ fermions, $N_{\Phi} = 21$, $N_A = 4$ and $l_A=11$. A small system-size has been selected for pedagogical purposes. The low-lying part of these spectra are almost identical, and exhibit the same structure and state-count. In addition to the Laughlin-like branch starting at $L_{z,A}=24$, the Coulomb spectrum contains at least two other clearly-defined branches starting at $L_{z,A}=28$ and $L_{z,A}=30$.}
\end{figure}

Figure \ref{figure1} shows a typical OES at $\nu=1/3$ for both the Laughlin state (fig. \ref{figure1}.a) and the Coulomb interaction ground state (fig. \ref{figure1}.b). In the Laughlin state, the entanglement stops at a maximum value of $L_{z,A}=L^{max}_{z,A}$, the $z$ component the angular momentum of the complementary region B. At this value, a single state is found for any values of the  total particle number, despite the fact that the corresponding Hilbert space dimension of states in $A$ grows exponentially with $N_A$, the number of particles in the sector investigated.  When applied to the Laughlin state root configuration, the orbital partitioning results in the root configuration $10010010010$ for the region $A$, which corresponds to the single state found at the highest possible value of $L_{z,A}$. While the existence of a root partition makes it obvious that no state can be found with a higher $L_{z,A}$, the fact that there is a unique state for this value is nontrivial and is related to a strong constraint on the state decomposition on the Fock basis called the "product rule"\cite{bernevig-09prl206801,thomale-2010arXiv1010.4837T}. The coefficient of a configuration whose two parts after a cut can be independently-obtained  by squeezing from the state root partition  is equal to  the product of the coefficients of the two disconnected pieces. When decreasing $L_{z,A}$ from this value, the counting of the eigenvalues  (the number of eigenvalues at a given $L_{z,A}$)  of the model state is the same  as the number of levels of the corresponding edge conformal field theory (CFT)  in the thermodynamic limit;  empirically, it consists of counting the number of $(1,3)$-admissible partitions with the correct angular momentum in $l_A + \Delta L_z$ orbitals, where $\Delta L_z = L^{max}_{z,A}-L_{z,A}$. This conjecture can be proved using bulk-edge correspondence\cite{2011arXiv1102.2218C}. In the finite size systems, the equivalence between thermodynamic CFT edge counting and orbital entanglement spectrum counting does not hold for all $L_{z,A}$ values. The counting of the spectrum develops finite-size effects that have been recently \cite{hermanns-1009arXiv4199H} related to the encoding of the Haldane exclusion principle within the model state.

The entanglement spectrum of the Coulomb interaction ground state at filling $\nu=1/3$ exhibits a branch of low-lying levels displaying the same CFT counting as the Laughlin state, separated from higher energy states by an entanglement gap which was conjectured to remain finite in the thermodynamic limit\cite{li-08prl010504}. However, this entanglement gap closes as $L_{z,A}$ is reduced, making its definition ambiguous. Using a different Fock state normalization in which each LL orbital is normalized by the same factor, it has been shown that, in the Coulomb ground state OES, a full gap emerges between low-lying states whose state-count is the same as that of the Laughlin state, and higher entanglement energy states\cite{thomale-10pr180502} which had been previously deemed non-universal. In this paper (in section \ref{section_laughlinCoulombES})  we show that even the higher entanglement energy states in fact present a universal structure  related to particle-hole excitations of the Haldane pseudopotential Hamiltonian for the Laughlin state. For now, we notice in the Coulomb OES the presence of well separated structures ("christmas-tree"-like branches)  with states starting at a larger $L^{max}_{z,A}$ than that of the Laughlin-like branch (for example, the branches starting at $L^{max}_{z,A}=28$ and $L^{max}_{z,A}=30$ in Figure \ref{figure1}.b).   Understanding these branches is the purpose of this paper.

\section{Composite fermions wavefunctions and their excitations}\label{section_JainCF}

\begin{figure}[htb]
\centering
\includegraphics[width=9 cm]{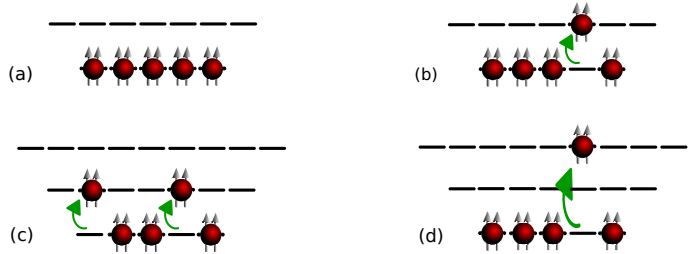}
\caption{\label{fig_cf}(a) Schematic representation of the Laughlin state as a CF filled lowest $\Lambda$ level for $N=5$ CFs. (b) Creation of a quasiparticle-quasihole excitation above the Laughlin state in the  $L_z=0$ sector. (c) and (d) Two different possibilities to create an excited state with energy $2\hbar\omega^*_c$.}
\end{figure}

Jain's ``composite fermion'' (CF) picture \cite{jain89prl199} provides a nice heuristic explanation of many features of quantum Hall effect, including the observed incompressible states at $\nu=p/(2p+1)$ as well as  the  existence of a compressible the state at $\nu=1/2$ (see reference \cite{Jain_CF} for an extensive review on CF). The CF ansatz replaces the strongly interacting electrons or bosons by   ``composite fermions''  formed by  binding them to $n$ flux quanta, where  $n$ is even or odd,  depending on whether the original ``bare''  particles are bosons ($n$ odd) or fermions ($n$ even). The generic Jain states are given by:
\begin{equation}
\Psi_{CF}={\cal P}_{{\rm LLL}} \left[\prod_{i<j}\left(u_iv_j - u_jv_i\right)^n \Phi^{CF}_p\right]
\label{jaincfwf}
\end{equation}
\noindent where $\prod_{i<j}(u_iv_j - u_jv_i)^n$ binds  $n$ flux quanta to each original particle. ${\cal P}_{{\rm LLL}}$ is the projection operator onto the LLL. $\Phi^{CF}_p$ is the wave function of the free CF in $p$ effective Landau levels, which Jain has called ``Lambda levels'' ($\Lambda$L).  When $p$ such levels are fully occupied, equation \ref{jaincfwf} gives rise to a model state that remarkably accurately approximates the ground state at filling $\nu=p/(np+1)$.  The observed incompressible states at $\nu=p/(2p+1)$ can then  be  thought of  as  incompressible integer QH CF states at $\nu^*=p$ where $\nu^*$ is the filling factor of the CFs. Moreover, the compressibility of the $\nu=1/2$ state can be attributed\cite{halperin-PhysRevB.47.7312} to the formation of a CF Fermi sea state as the effective magnetic field felt by the CFs vanishes. In this picture, the $\nu=1/m$ Laughlin wave function is interpreted  as the limiting case  $n=m-1$ with $p=1$ filled CF $\Lambda$ levels. A schematic view of this picture of the Laughlin state is shown in figure \ref{fig_cf}a.  Except for
the Laughlin state, the  Jain wavefunctions are not known to be unique zero-energy states of a model Hamiltonian.
However, it has been recently been shown that they do have some special ``squeezing'' properties:  the bosonic Jain states at $\nu=p/(p+1)$ and the bosonic counterpart $\Psi_B=\Psi_F/(\prod_{i<j}(u_iv_j - u_jv_i))$ of the fermionic one at $\nu=p/(2p+1)$ have a single root partition and vanish with a power 2 when $p+1$ particles are brought to the same point \cite{regnault-09prl016801}.

The natural way to build quasihole-quasiparticle excitations within the CF construction is to assume that the $\Lambda$L's are separated by an effective cyclotron energy $\hbar\omega^*_c$ and to then sort the different excited states with respect to their energy. Thus, the lowest energy excited states above the Laughlin ground-state, with energy is $\hbar\omega^*_c$, are obtained by exciting one CF in the second $\Lambda$L as shown in figure \ref{fig_cf}b. For $2\hbar\omega^*_c$ energy states, there are two possibilities: two CFs can be put in the second $\Lambda$L (fig. \ref{fig_cf}c) or one CF in the third $\Lambda$L  - all  other CFs remains in the lowest one (fig. \ref{fig_cf}c). In general, the $n^{th}$ excited states branch involves the $n+1$ lowest $\Lambda$ levels. Although this method is just a simple phenomenological sketch (as interactions between ``composite fermions'' should surely also be taken into account, and there is no phenomenological reason for equal spacing of the levels),  the resulting wavefunctions  remarkably-well reproduce the low energy structure of the Coulomb-interaction Hamiltonian when their variational energies are evaluated using the exact Hamiltonian and CF diagonalization\cite{Jain_CF}. 

Monte Carlo methods have been used extensively to quantitatively explore the predictions of the CF picture\cite{jain-97ijmpb156404}, proving useful in computing quantities such as predicted energies of CF states and their overlaps with states obtained by exact diagonalization.    However, calculations of entanglement spectra requires a high accuracy in the decomposition of the wave function on the Fock basis. With the exception of several special cases \cite{regnault-09prl016801}, Monte Carlo techniques fail to reach the accuracy needed and we needed  to implement a new exact projection method.  The first step of this method is to use the Jack recursion formulas of \cite{bernevig-09prl206801,stanley89advm76} to expand $\prod_{j<k}(u_jv_k-v_ju_k)^n$ (which is a Jack polynomial)  in terms of symmetric monomials or slater determinants (for  $n$ even or odd respectively). The product with $\Phi^{CF}_p$ is computed explicitly, using only the resulting wave function symmetry properties to reduce the computation time.  Finally, the projection is applied and only terms fully in the LLL are kept. We found that the clustering properties characterizing how ``composite particle'' quasihole-quasiparticle excitation states vanish as particles come together are determined by the number of the $\Lambda$ level involved: the bosonic states and the bosonic counterpart of the fermionic states (fermionic states divided by a vandermonde determinant) constructed from the $k$'th  $\Lambda$L vanish when $k+1$ particles are brought to the same point as the second power of the difference in the particle coordinates. Therefore the excited states also exhibit a non-trivial root configuration. In all the cases we have studied, we noticed the following properties of states of energies less or equal to $2\hbar\omega^*_c$: such excited states can be realized in a number of different ways:  all CFs in the lowest $\Lambda$L, one or two CFs in the second $\Lambda$L or one CF in the third $\Lambda$L. The last possibility does not introduce any new states compared to the ones that are produced using the first two ways. To obtain this result, the LLL projection is crucial as it creates  linear dependencies between the projected states and greatly reduces  the number of independent states. For instance, for $N=6$ fermions and $N_{\Phi}=15$ in the $L_z^{tot}=0$ sector, there are $51$ states with $2$ CFs or less in the second $\Lambda$L and none in the third LL whereas there are $57$ states whose energy is less than or equal to $2\hbar\omega^*_c$; after projection and re-orthonormalization these two numbers are reduced to $36$ and the states spaces spanned by these two sets of states are identical. 

\section{From Laughlin to Coulomb entanglement spectrum}\label{section_laughlinCoulombES}

In this section, we use the admixture of virtual quasiparticle-quasihole excited states into the ground state  to interpolate between the Laughlin state and the Coulomb interaction ground state at the same filling factor,  and  examine  the effect of this on the orbital entanglement spectrum. As both states have $L=0$ we only consider the zero angular momentum sector . The $\nu=1/3$ Laughlin state is known to be the densest zero-energy eigenstate  of the Haldane pseudopotentials\cite{haldane83prl605} by $V_1=1$ and $V_{n>1}=0$. Using this interaction, we could obtain the low energy $L=0$ states and use them to reconstruct the Coulomb interaction ground state. However, these states do not have an exact model wavefunction construction that would lead to a simple understanding of the additional structure in the orbital entanglement spectrum of the Coulomb state.  In particular the $L=0$ excited states of this model interaction do not feature a non trivial root configuration, contrary to the CF approach. Therefore, we first use the CF quasihole-quasiparticle excitations states whose construction was described in the previous section to iteratively construct a basis of $L=0$ states (the iteration being the number of $\Lambda$L levels  occupied) whose effective cyclotron energy is less than a given effective cyclotron energy. We approximate the Coulomb ground state using: 
\begin{equation}
\ket{\Psi_{n}}= \sum_{\ket{\Psi^{n\hbar\omega^*_c}_{CF}}} \braket{\Psi^{n\hbar\omega^*_c}_{CF}}{\Psi_{coulomb}}\ket{\Psi^{n\hbar\omega^*_c}_{CF}}
\label{Coulombfit}
\end{equation} 
where the sum runs over all the $L=0$ states of the iterative basis which spans the space of states whose effective cyclotron energy is less than $n\hbar\omega^*_c$. This state is then normalized, which translates into a global shift of the entanglement spectrum. As the effective cyclotron energy is increased, the partition that dominates all the basis states changes and  successively dominates the previous basis (see Table \ref{table_cf}). The $L^{max}_{z,A}$'s of the successive root partitions  correspond to those of the additional structures observed on Coulomb ground state entanglement spectrum - thereby allowing us to predict the starting point of the higher entanglement energy branches. In the Table \ref{table_cf}, the number of $L=0$ states, the topmost partitions and the corresponding $L^{max}_{z,A}$ values for the accessible effective cyclotron energy are given for $N=8$ and $N_{\Phi}=21$.
 
%For each numerically accessible value of this energy, \ie $2,3,4 \hbar\omega^*_c$, as there is no $L=0$ state with an energy of $\hbar\omega^*_c$ and that $5\hbar\omega^*_c$ is not feasible for the moment.

\begin{table}[htb]
\begin{center}
\begin{tabular}{c|c|c|c}
$n$ & number of $L=0$ states & Topmost partition &$L^{max}_{z,A}$\\
\hline
1&1&10010010010\textbar01001001001&24\\
2&4&11000100100\textbar00100100011&28\\
3&8&11000100100\textbar01000010011&28\\
4&14&11001001000\textbar00010010011&30\\
\end{tabular}
\caption{\label{table_cf} Characteristics of the iterative basis which spans the space of states whose effective cyclotron energy is less than $n\hbar\omega^*_c$ for $N=8$ and $N_{\Phi}=21$. The last column indicates the $L^{max}_{z,A}$ value corresponding to the topmost partition with respect to the cut defined by $L_A=11$ and $N_A=4$ -  shown by a vertical line in the partitions. The total Hilbert space has $31$ $L=0$ states - we are using at most $14$ of them.}
\end{center}
\end{table}

\begin{figure}[htb]
\includegraphics[width=5.3 cm]{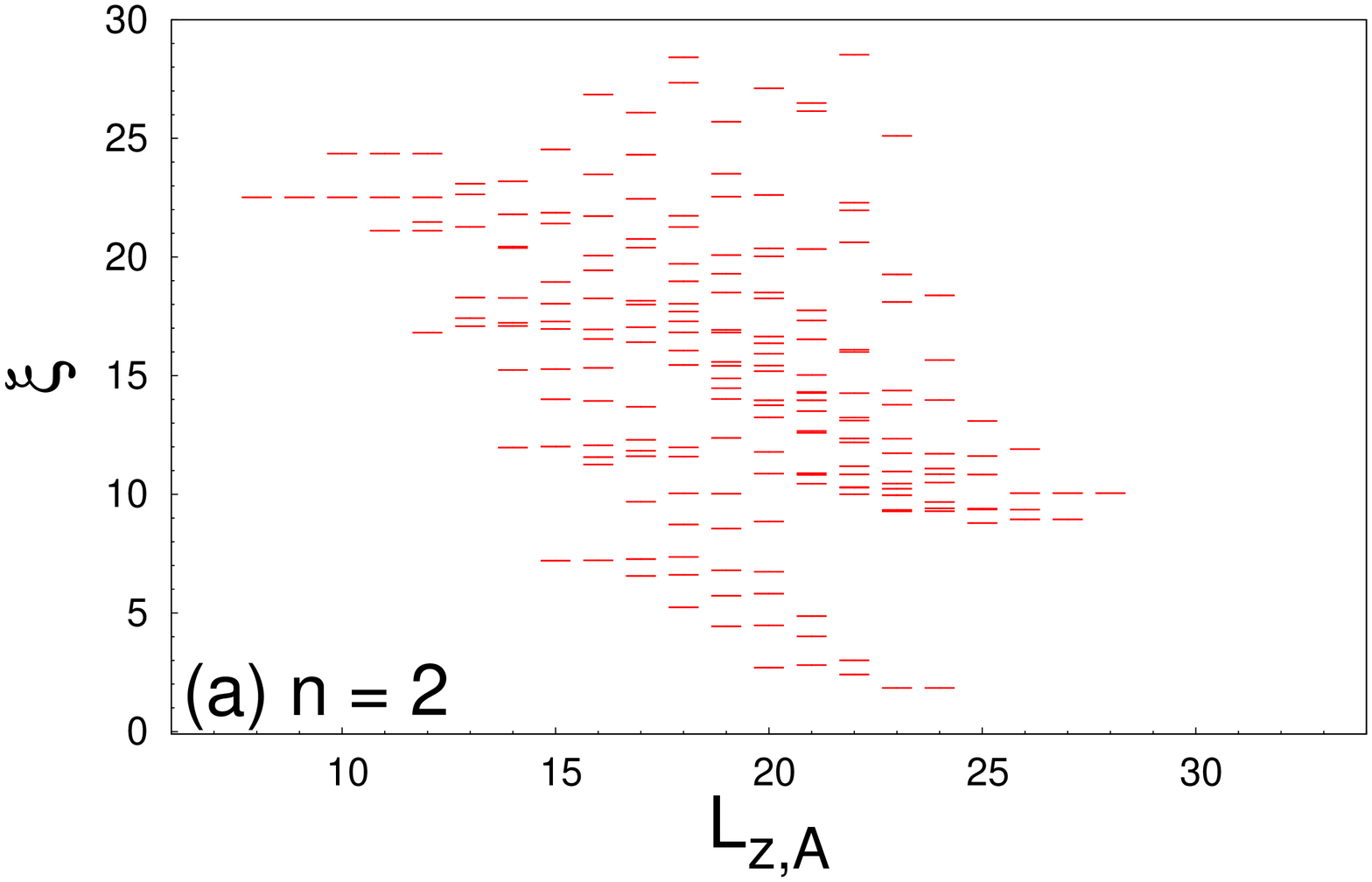}
\includegraphics[width=5.3 cm]{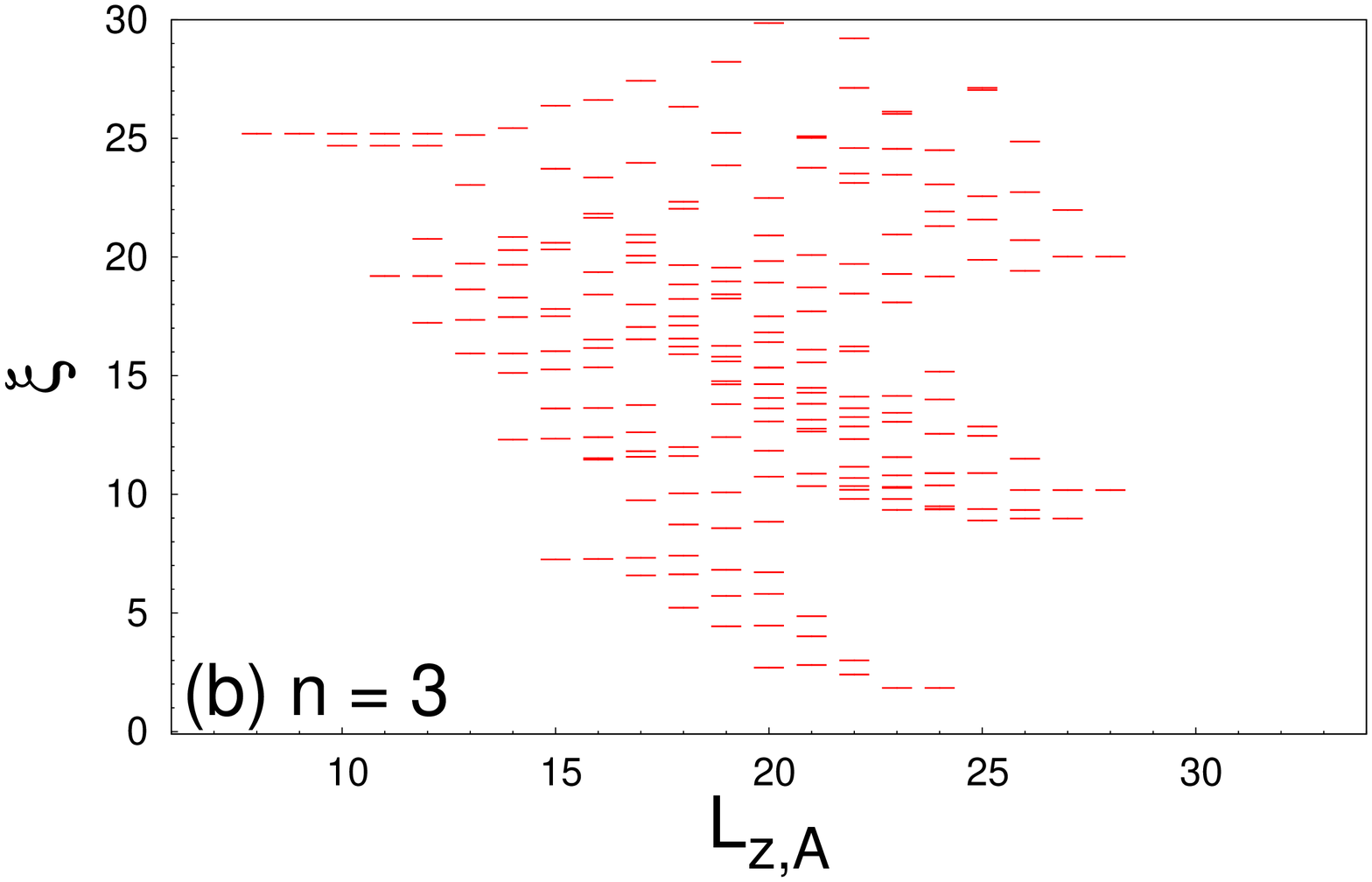}
\includegraphics[width=5.3 cm]{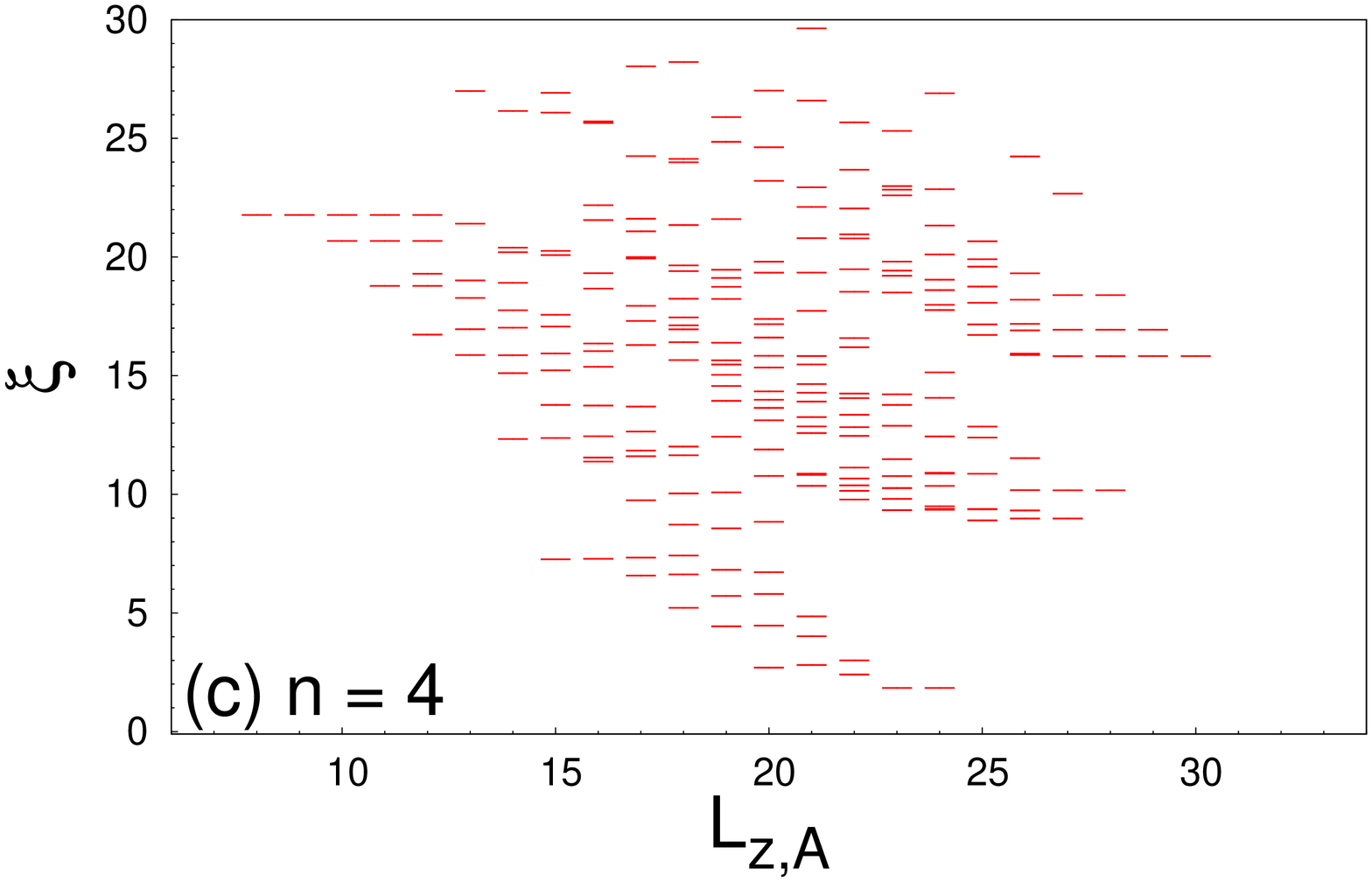}
\caption{\label{oes_cf} Orbital entanglement spectrum of different $\ket{\Psi_{n}}$: $n=2$ (a), $n=3$ (b) and $n=4$ (c) for $N=8$ fermions, $N_{\Phi} = 21$, $N_A = 4$ and $l_A=11$. The structures observed on the Coulomb state OES progressively appear as $n$ is increased and we obtained completely seperated branches when $n$ is even. The last observed branch in the Coulomb spectrum, for which $L^{max}_{z,A}=32$,  is expected to be described by an higher effective cyclotron energy CF  wavefunctions.}
\end{figure}

\begin{figure}[htb]
\includegraphics[width=5.3 cm]{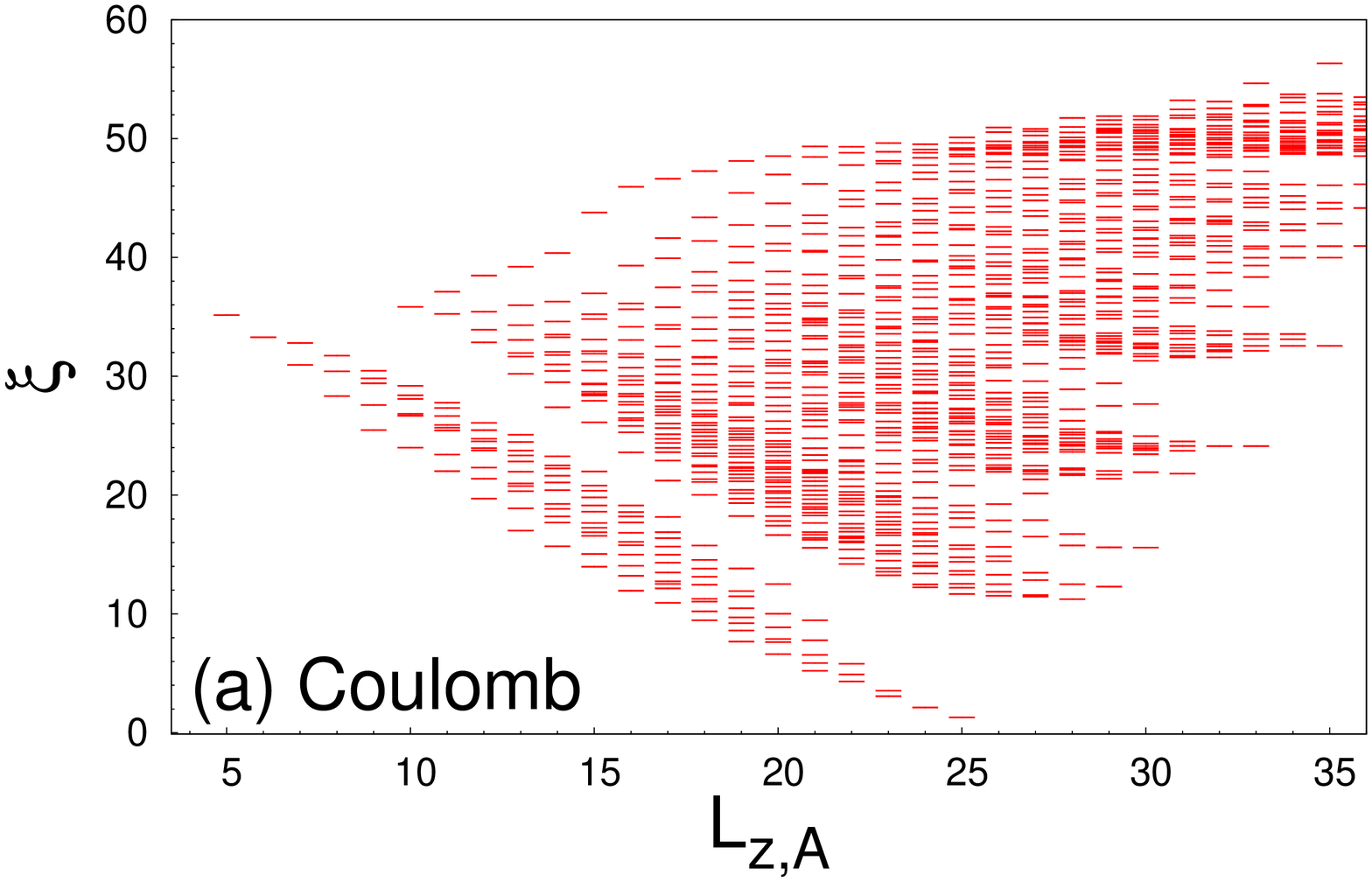}
\includegraphics[width=5.3 cm]{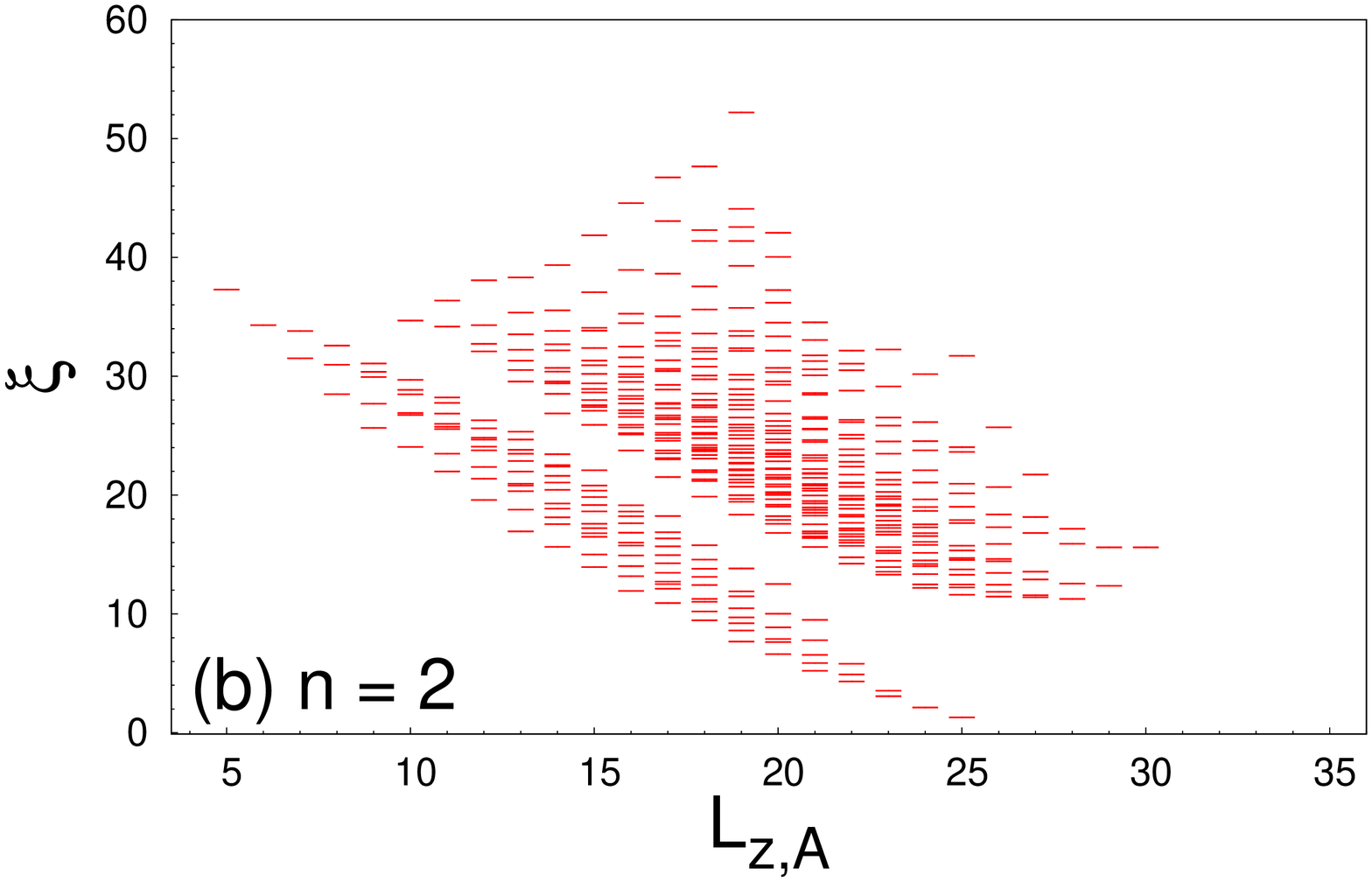}
\includegraphics[width=5.3 cm]{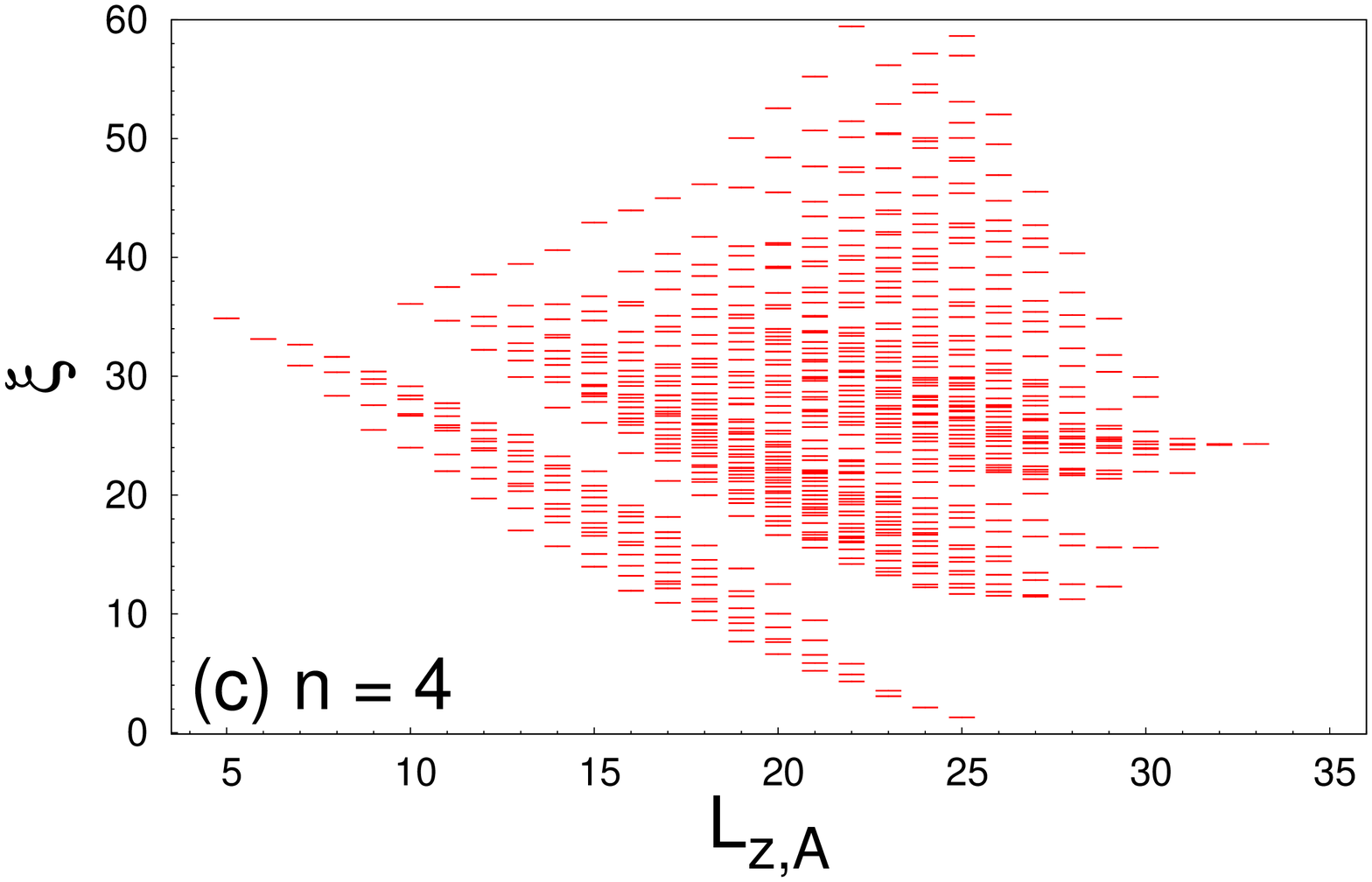}
\caption{\label{oes_cf_bosons} Orbital entanglement spectrum of for the Coulomb interaction ground state (a) and different $\ket{\Psi_{n}}$: $n=2$ (b) and $n=4$ (c) for $N=10$ bosons, $N_{\Phi} = 18$, $N_A = 5$ and $l_A=9$.}
\end{figure}

The comparison of the orbital entanglement spectrum for the first three interesting $\ket{\Psi_{n}}$ displayed in Figure \ref{oes_cf}, and the Coulomb state spectrum (Fig. \ref{figure1}) shows that the  CF quasihole-quasiparticle excitation basis allows a step-by-step reconstruction of the different Coulomb entanglement spectrum structures. Given the method used, it is obvious that the spectra of $\ket{\Psi_{n}}$ and $ \ket{\Psi_{coulomb}}$ would get closer as the effective cyclotron energy is increased; in that case, we include more and more $\Lambda$L's and, 
in a  finite-size calculation, we start  diagonalizing in larger and larger parts of the full Hilbert space. However, nontrivially, the structures of the Coulomb ground state entanglement spectrum  are  sorted by the effective cyclotron energy, in the sense that the first, second, \textit{etc.} branches  of the Coulomb spectrum can be obtained by only using CF states with $n=1,2$, \textit{etc}. We checked these properties for the fermionic $\nu=1/3$ Laughlin state for up to $N = 8$ particles and for the bosonic $\nu = 1/2$ one for up to $N = 10$ particles (Fig. \ref{oes_cf_bosons}). The different scales in the eigenvalues of the Coulomb interaction ground state density matrix seem to be  linked to the different effective cyclotron energy scales involved. 

\section{Orbital entanglement spectrum at finite temperature}\label{section_thermalES}

\begin{figure}[htb]
\begin{center}
\includegraphics[width = 5.1 cm]{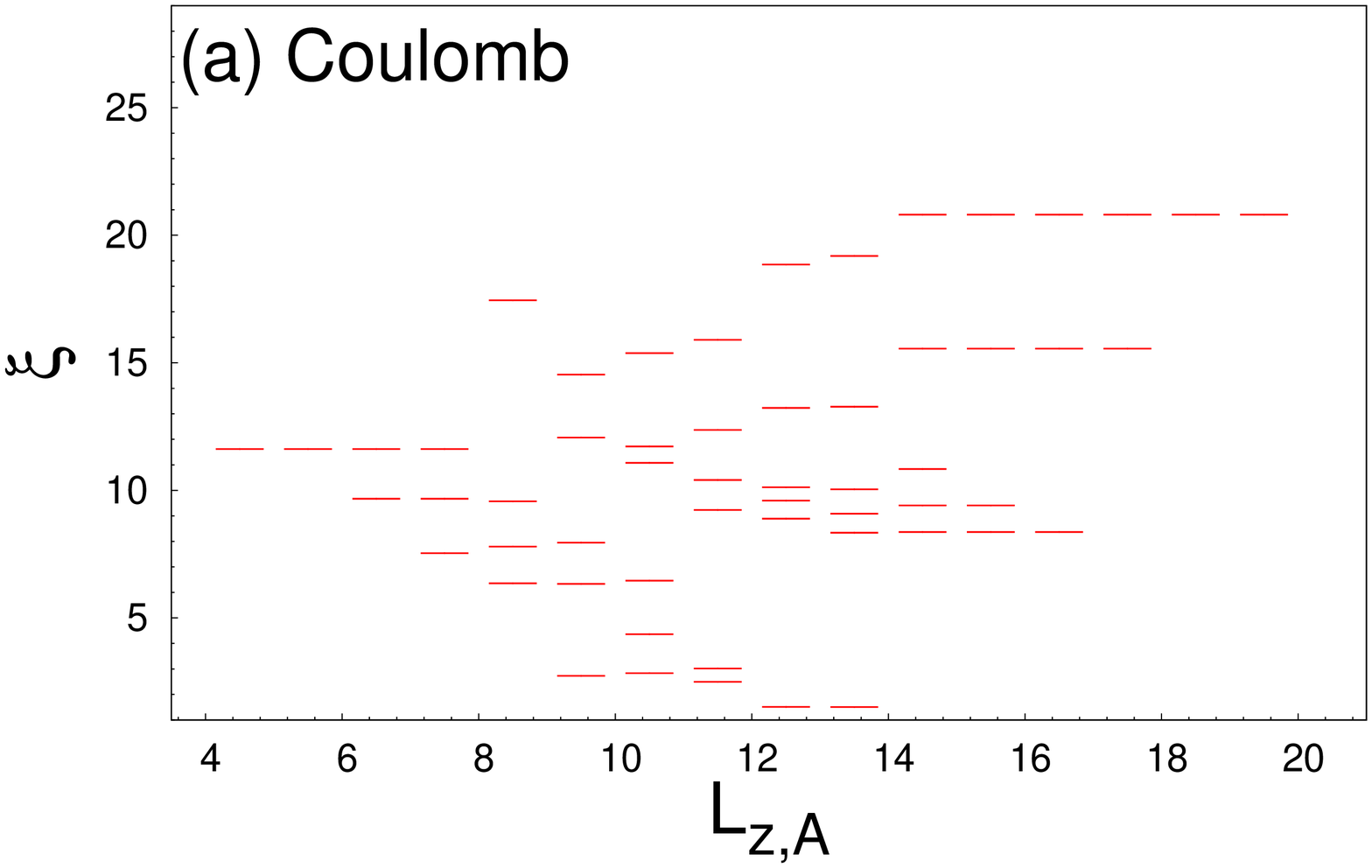}
\includegraphics[width = 5.1 cm]{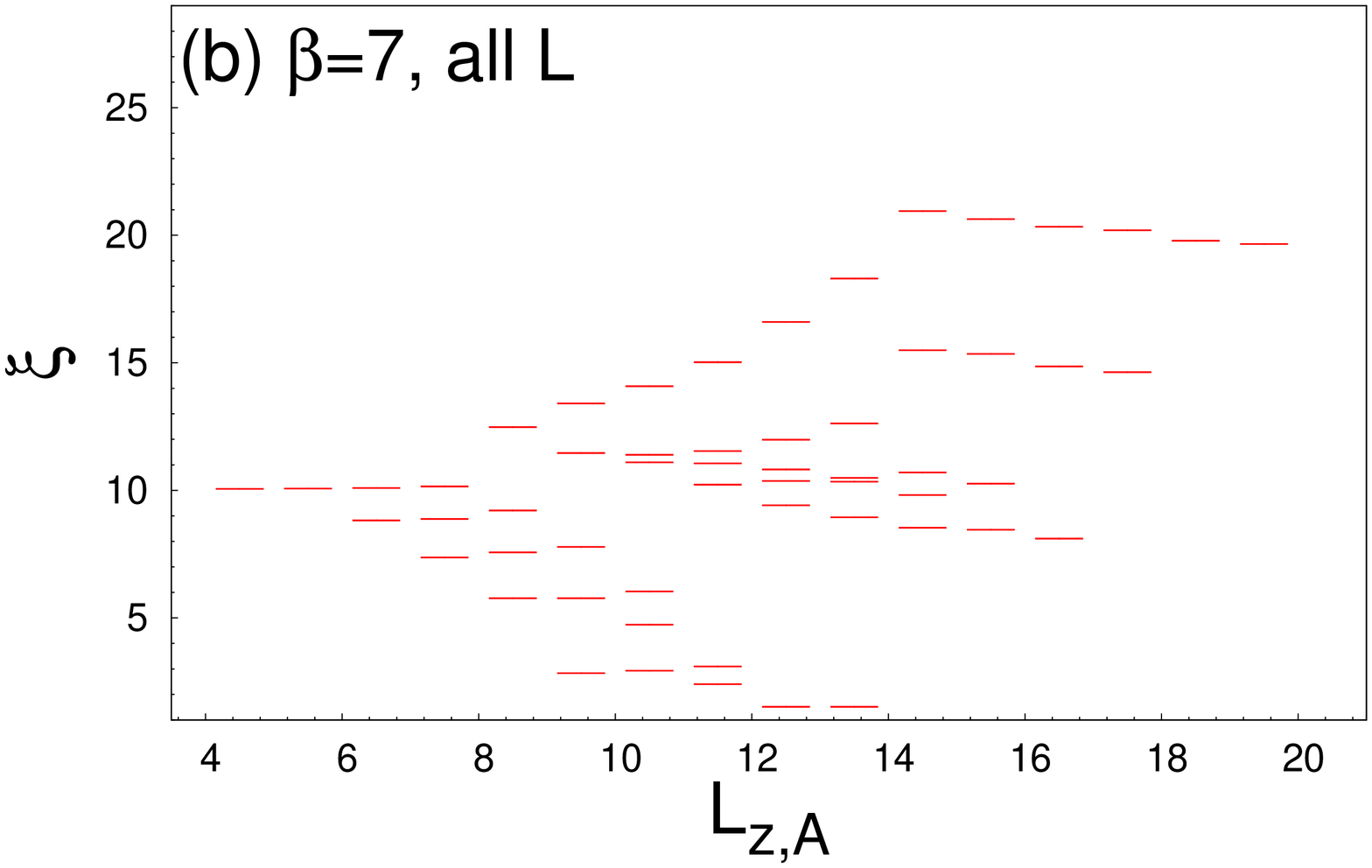}
\includegraphics[width = 5.1 cm]{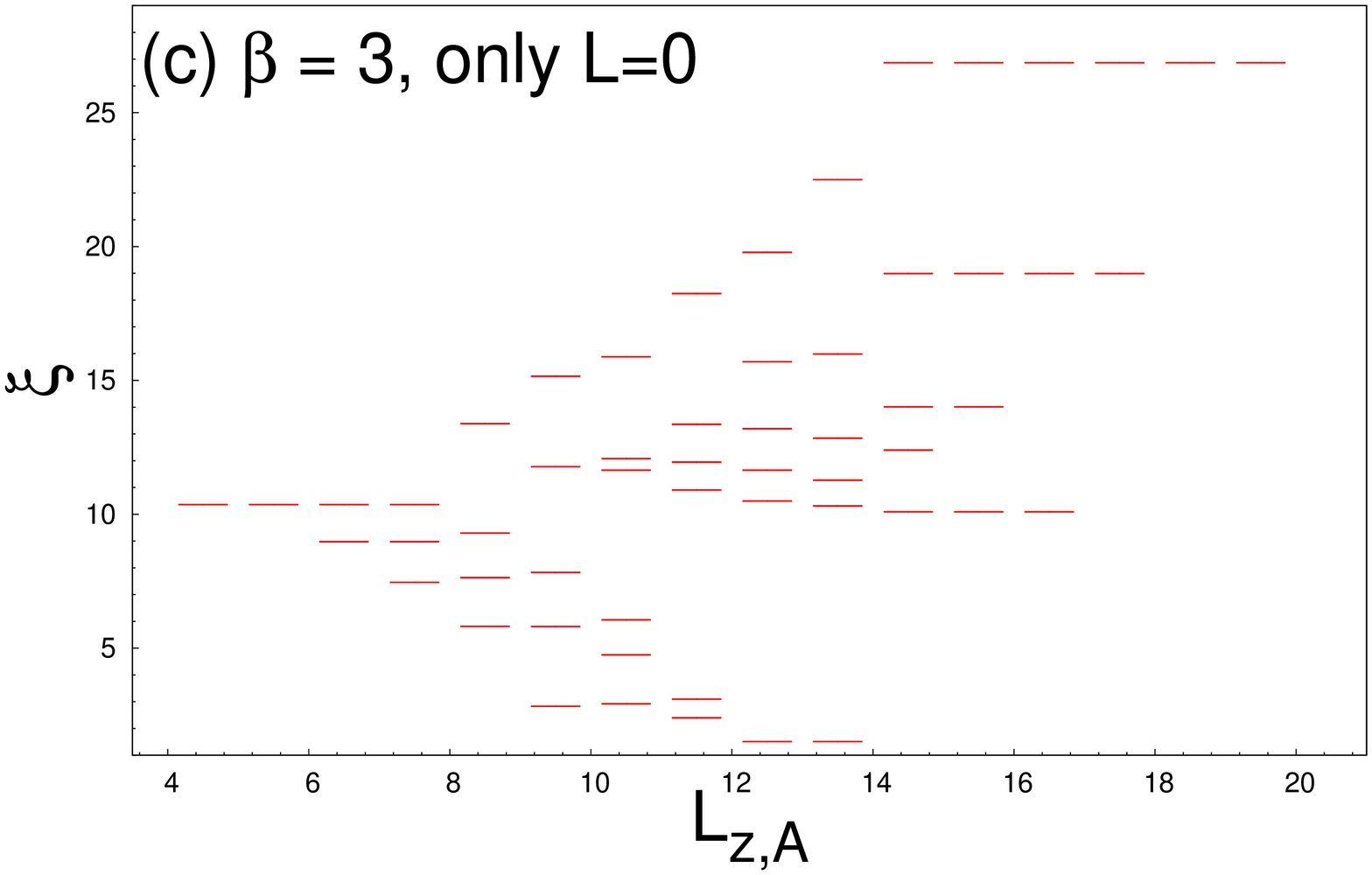}
\caption{\label{fullspectrum} Orbital entanglement spectrum for the Coulomb ground state (a), for $\rho$ defined by equation \ref{thermaldens} with $\beta=7$ (b) and $\rho'$ defined by equation \ref{thermaldens} but with the sum restricted to the $L=0$ states with $\beta=3$ (c) for $N=6$, $N_{\phi}=15$, $N_A=3$ and $L_A=8$. The value of the inverse temperature $\beta$ has been chosen so that the spectra are similar to that of the Coulomb ground state. This naturally leads to a choice of  two different temperature values for $\rho$ and $\rho'$ as the number of states considered in these two cases is very different. The general shape of these spectra is very similar.}
\end{center}
\end{figure}

We now try to reproduce the Coulomb ground state entanglement spectrum using a thermal density matrix. For a system at finite temperature $T$, the density matrix is given by:
\begin{equation}
\rho = \frac{1}{Z}\exp(-\beta H)\label{thermaldens}
\end{equation}
where $\beta=1/T$, $Z=\Tr\left[\exp(-\beta H)\right]$ and $H$ denotes the Hamiltonian of the system. In our case, we want to compare the entanglement spectrum of the Coulomb-interaction ground state to the entanglement spectrum of  the density  matrix describing finite-temperature  corrections to the Laughlin state and hence the simplest choice is to take $H$ to be the pseudopotential interaction Hamiltonian. At $T=0$ we recover the pure  Laughlin-state entanglement results. In contrast to the previous section, we now have to take into account not only the $L=0$  states, but states with all possible  values of $L$, and all the $L_z$ sectors of the pseudopotential Hamiltonian. This involves a very large number of states and  Equation \ref{thermaldens} can be realized exactly only for small systems. By choosing an \textit{ad-hoc} "entanglement temperature" temperature $T$, the entanglement spectrum of $\rho$ and the one of the Coulomb ground state can be made very similar as shown on figure \ref{fullspectrum}. Although the full shape of the spectrum is very similar, we notice that the Coulomb spectrum exhibits degeneracies which are not  present in the thermal density matrix approach. For example, the rightmost levels are degenerate for the Coulomb-interaction ground state and are no longer degenerate when the thermal density matrix is used, despite its rotational-invariance. It can be analytically shown that these degeneracies are linked to  the $L = 0$   states that enter the density matrix.  When the sum of Equation \ref{thermaldens} is restricted to just the $L=0$ states, these degeneracies are recovered (fig.  \ref{fullspectrum}c).

\begin{figure}[htb]
\begin{center}
\includegraphics[width = 7.5 cm]{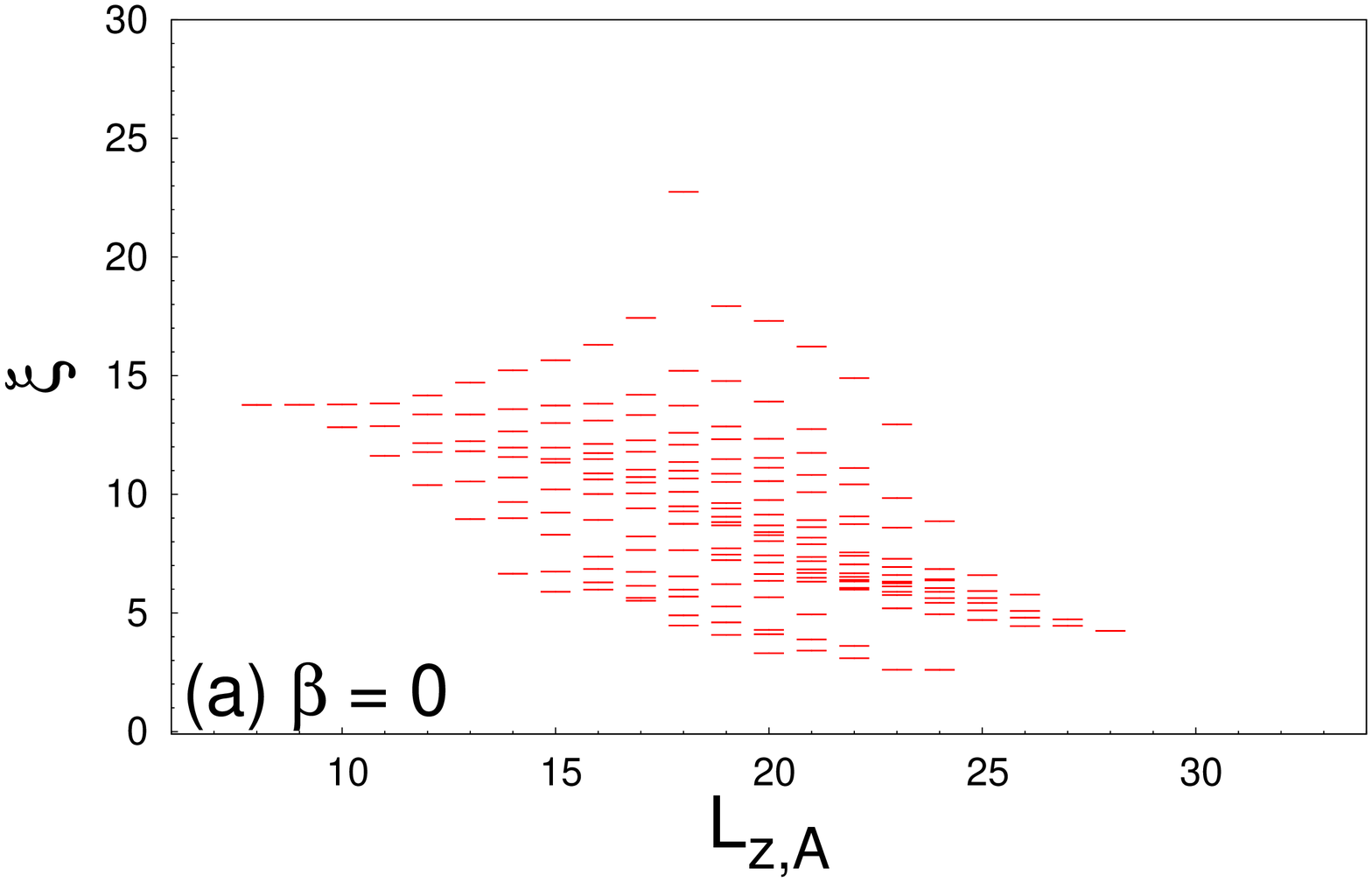}
\includegraphics[width = 7.5 cm]{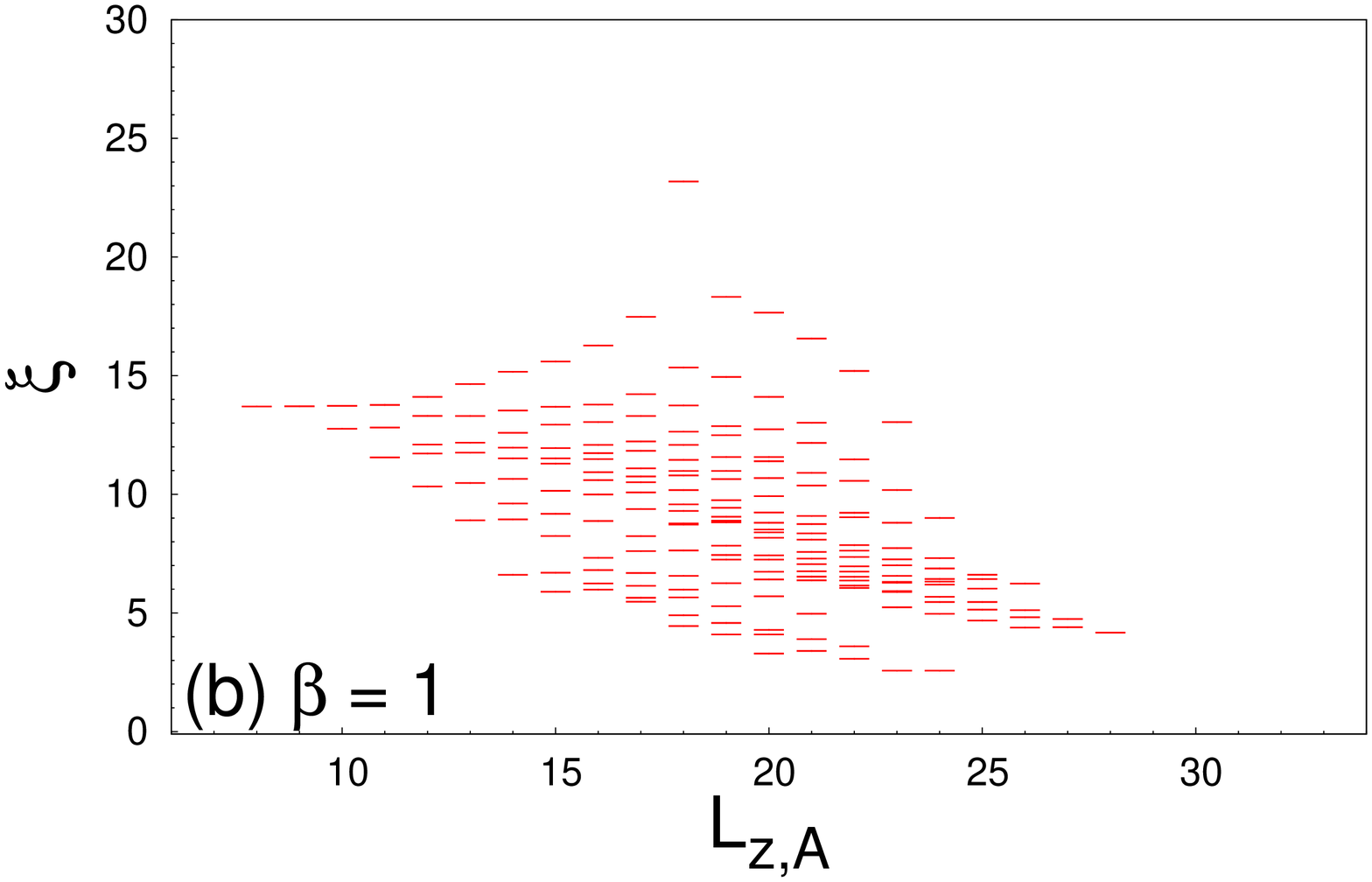}\\
\includegraphics[width = 7.5 cm]{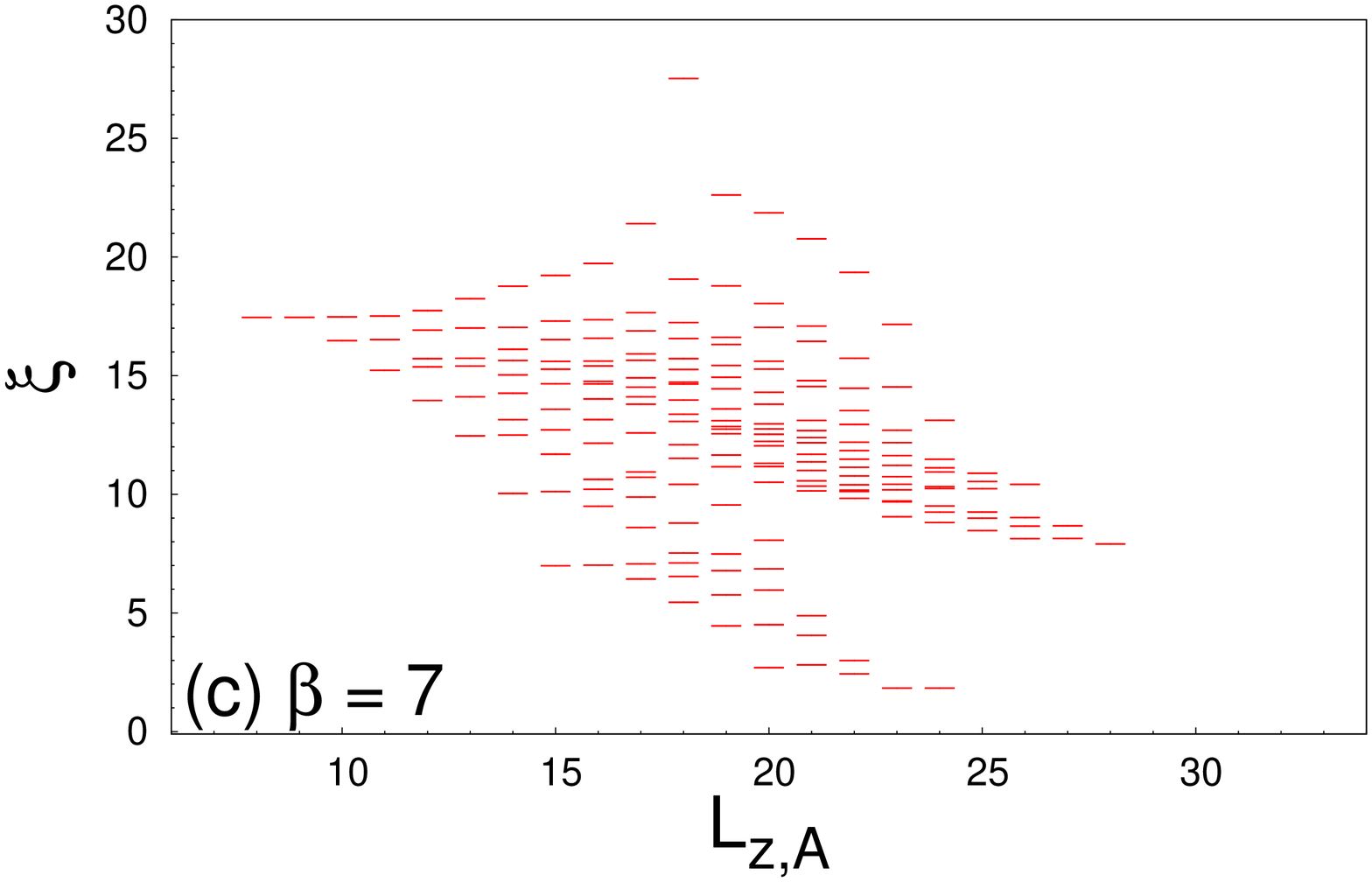}
\includegraphics[width = 7.5 cm]{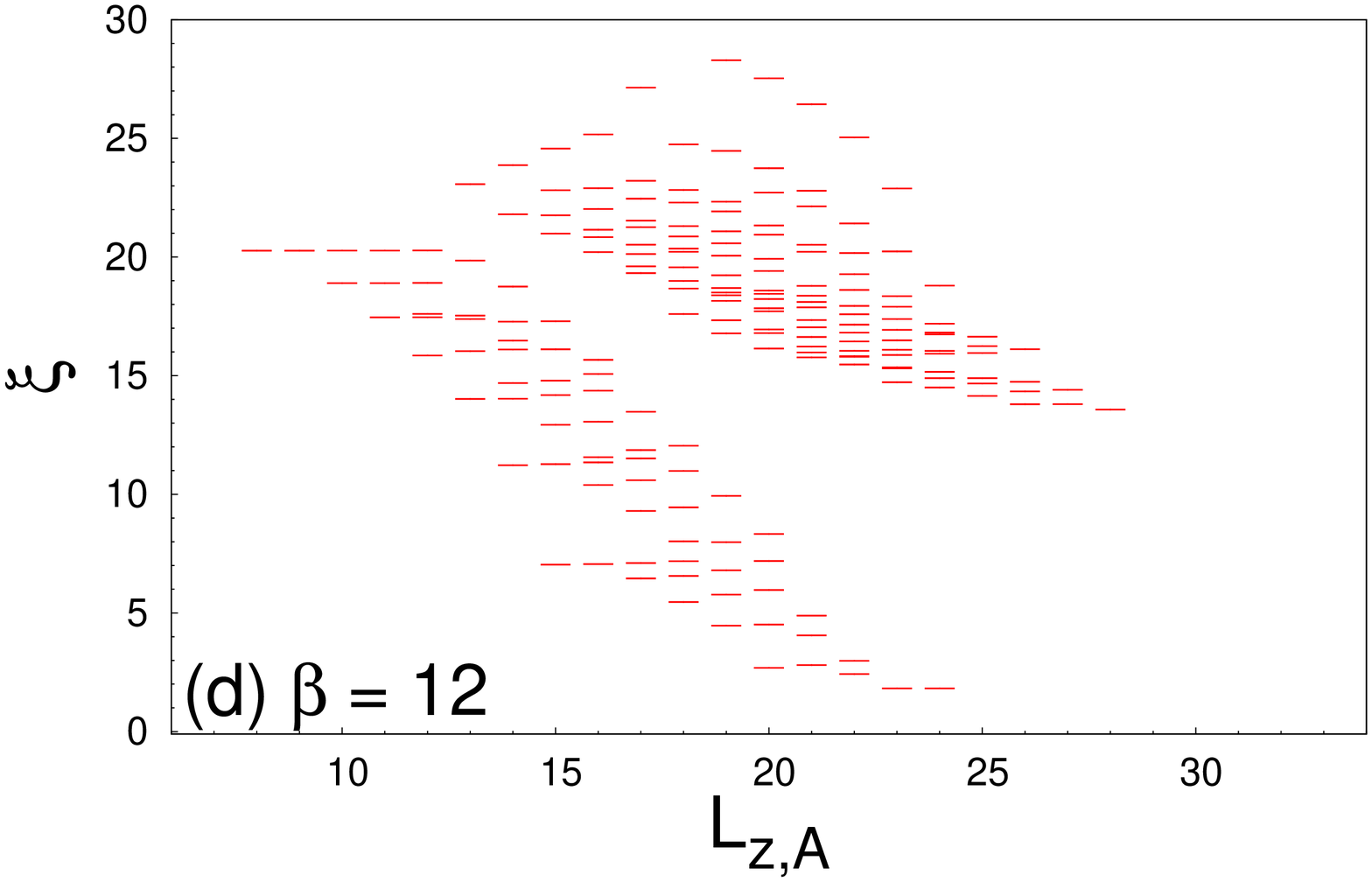}
\caption{\label{ESMagn}Entanglement spectrum of the single-mode approximation $\rho$ given by equation \ref{eqmagndens} for a system of $N=8$ fermions and $N_{\Phi}=21$ flux quanta, (a): $\beta=0$, (b): $\beta=1$, (c): $\beta=7$, (d): $\beta=12$. The cut used is $N_A=4$ and $l_A=11$.}
\end{center}
\end{figure}

To analyze bigger systems, we consider a simpler model. In the thermal density matrix, we only kept the two lowest energy branches of the incompressible state: the ground state wave function and the magneto-roton mode. This is the single-mode approximation of the entanglement spectrum. The dispersion relation of the magneto-roton mode is that obtained through by the $V_1$ interaction. The magneto-roton states $\ket{\Psi^{mag}_{L,lz}}$, obtained by putting one CF in the second $\Lambda$L, consist of $N-1$ multiplets whose total angular momentum ranges from $L=2$ to $L=N$. Even though a $L=1$ state is naively expected, this multiplet is systematically suppressed by the $LLL$ projection, as  pointed out in \cite{PhysRevLett.69.2843}. Thus we use the following approximation for $\rho$:
\begin{equation}
\label{eqmagndens}\rho=\frac{1}{Z} \left(\ket{\Psi_{Laugh}}\bra{\Psi_{Laugh}} + \sum_{L,l_z} e^{-\beta E_L} \ket{\Psi^{mag}_{L,lz}} \bra{\Psi^{mag}_{L,lz}}\right)
\end{equation}
where $$Z=1 +\sum_{L,l_z}e^{-\beta E_L}.$$ $L$ is the total angular momentum of the multiplet states and $E_L$ is their energy with respect to the pseudopotential interaction Hamiltonian. We then compute the orbital entanglement spectrum associated with this density matrix for various $\beta$ values and for up to $N=10$ fermions (see Fig. \ref{ESMagn}).

In every case we have studied, when the temperature is infinite, the universal Laughlin-like part of the entanglement spectrum cannot be distinguished. As temperature is decreased, two branches of levels split: a lower one whose state-count is the same as that of the  Laughlin state, and an upper one whose average ``entanglement energy'' goes to infinity as $T$ reaches zero. In the sector $L_{z,A}=L^{max}_{z,A}$ (in Fig[\ref{ESMagn}] $L^{max}_{z,A}=24$),  and the neighboring ones, it is possible  even at infinite temperature (see figure \ref{ESMagn}),  to define an entanglement gap \cite{li-08prl010504},  given by $\delta_{L_{z,A}}=\xi_{n+1,L_{z,A}}-\xi_{n,L_{z,A}}$ where $\xi_{n,L_{z,A}}$ is highest entanglement energy at $L_{z,A}$ that belongs to the Laughlin-state structure. For example, in the infinite temperature limit in   figure \ref{ESMagn}, this can be done for $L_{z,A}=24,23,22$, at which the Laughlin-like levels of counting $1,1,2$ respectively are separated from the higher energy states by a small but visible gap.  Moreover, the upper-branch state-count, and the $L_{z,A}$ value at which it starts, are the same as those of the first branch after the Laughlin one in the Coulomb ground state entanglement spectrum. As the magneto-roton mode can be obtained from excitations to the first $\Lambda$L level, only the first entanglement spectrum above the Laughlin one can be fitted in this single-mode approximation.

\begin{figure}[htb]
\begin{center}
\includegraphics[width = 8 cm]{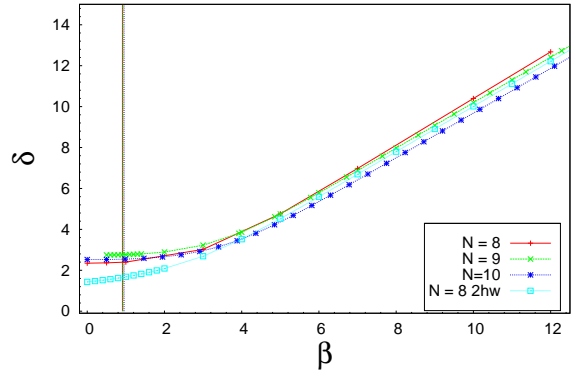}
\caption{\label{gapbeta}Entanglement gap $\delta_{L^{max}_{z,A}}$ for different temperature values and different number of particles: $N=8$ in red, $N=9$ in blue and $N=10$ in green and in turquoise $N=8$ with all the states whose effective cyclotron is less than $2\hbar\omega^*_c$. The vertical line is the temperature corresponding to the true energy gap.}
\end{center}
\end{figure}

\begin{figure}[htb]
\begin{center}
\includegraphics[width = 7.5 cm]{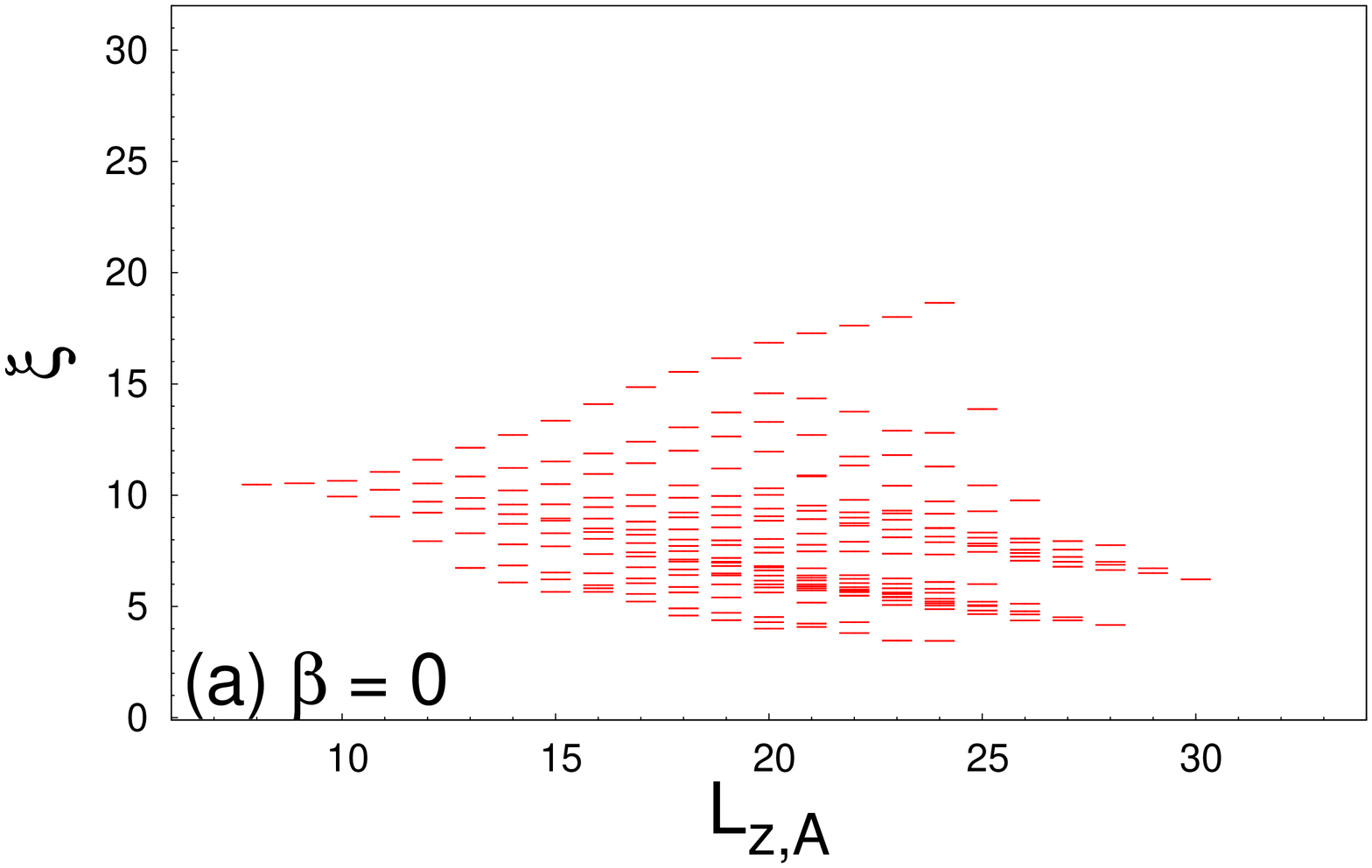}
\includegraphics[width = 7.5 cm]{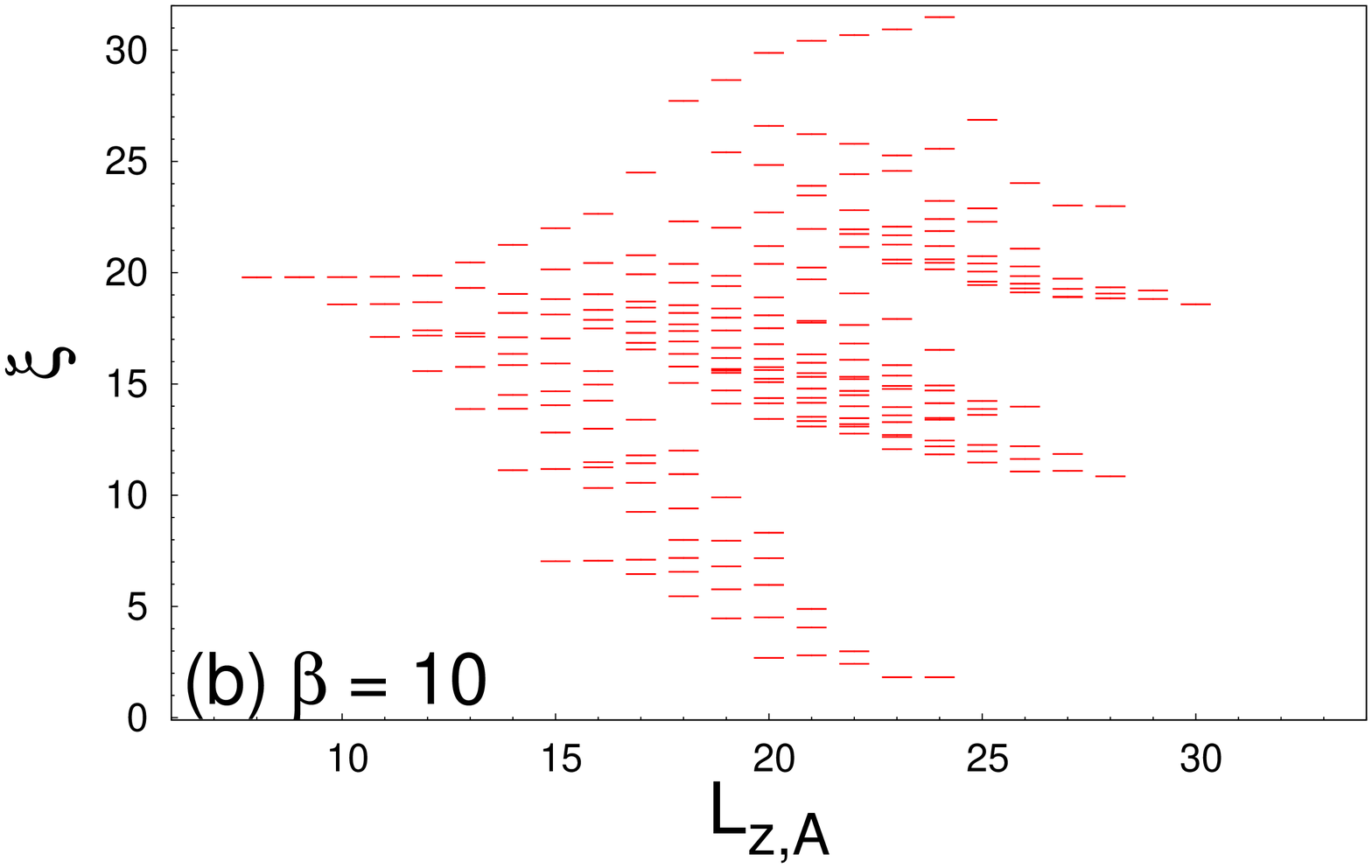}
\caption{\label{2cf}: Entanglement spectrum of $\rho$ given by a formula completely analogue to  \ref{eqmagndens} in which the $2\hbar\omega^*_c$ effective cyclotron energy states are taken into account, for a system of $N=8$ fermions and $N_{\Phi}=21$ flux quanta, (a): $\beta=0$ and (b): $\beta=10$. The cut used is $N_A=4$ and $l_A=11$.}
\end{center}
\end{figure}

To more precisely characterize  the behavior of the two branches as a function of the temperature,  we calculated the entanglement gap $\delta_{L^{max}_{z,A}}$ as a function of $\beta=1/T$. The entanglement gap, shown in Figure \ref{gapbeta}, decreases as the temperature is increased from $T=0$ but, starting from a certain temperature value of the order of the energy gap, it reaches a plateau. This behavior is the same for all the cases we studied, independent of the  number of fermions.   Moreover, to see how our crude single-mode approximation affects the entanglement spectrum, we considered the next energy-level feature, taking all CF excitations states whose effective cyclotron energies are less than or equal to  $2\hbar\omega^*_c$. These spectra are presented in Figure \ref{2cf}, where we observe the emergence of a third branch in the thermal density matrix, matching the third branch of the Coulomb entanglement spectrum. The main behavior as the temperature is varied is not substantially affected by these additional states. However, it should be noticed that an additional structure appears and that the entanglement gap at high temperature no longer exhibits a plateau but just a change of slope  (shown in the Figure \ref{gapbeta}). 

\section{Discussion}\label{section_discussion}

The entanglement spectrum can be used to determine the universality class of realistic Hamiltonians in a topologically ordered phase: by identifying the state-count of  the  lowest-lying entanglement branch  of the spectrum with the state-count (characters) of the CFT of an edge theory, we can in principle predict that the ground-state of the system lies in a certain topological phase.  In this paper we showed that, for the case of the $\nu=1/3$ Coulomb interaction, not only the low-lying branch (with the Laughlin-state state-count), but also the higher' ``entanglement-energy'' branches exhibit a nontrivial characteristic structure, related  to dressing of the simple model ground state that
gives rise to the lowest branch of the entanglement spectrum by zero-point fluctuations of particle-hole collective excitations, when the difference between the model Hamiltonian and the Coulomb Hamiltonian is added back as a ``perturbation'' .
We have explicitly showed that the  higher-energy branches  correspond  to the wavefunctions for adding particle-hole excitations on top of the Laughlin ground-state. The correspondence exhibits quantum number matching ($L_{z,A}$  and counting of the levels in a specific branch) with the Coulomb spectrum if the model wavefunctions we use for the excitations are Jain's CF wavefunctions. We then performed a single-mode approximation of the Coulomb spectrum by calculating the reduced thermal density matrix of the Laughlin state augmented by the magnetoroton mode. We found that by this method we could obtain the first branch above the Laughlin-state branch in the entanglement spectrum of the Coulomb state. This exercise also appears to supports the idea that the entanglement spectrum of the ground-state of a realistic Hamiltonian contains information not only about the universality class of the ground-state but also about its excitations, which in a generic Hamiltonian
(unlike in free fermion Hamiltonians FQH model Hamiltonians) will be represented in the ground state properties though zero-point fluctuations of collective modes.

\section*{Acknowledgment}

BAB was supported in part by the Alfred P. Sloan Foundation,  NSF CAREER DMR-095242,  and by NSF-MRSEC  DMR-0819860
at the Princeton Center for Complex Materials.   FDMH was supported by DOE grant DE-SC0002140.   BAB also wishes to thank Microsoft Station Q for generous hosting during the last stages of preparation of this work.

\clearpage
\section*{References}
\bibliographystyle{unsrt.bst}
\bibliography{CoulombES.bib}

\end{document}